\documentclass[prd,aps,twocolumn,superscriptaddress]{revtex4}
\usepackage{amsmath}
\usepackage{epsfig}
\usepackage{axodraw}

\newcommand{\insertfig}[2]{\mbox{\epsfxsize=#1cm \epsfbox{#2.eps}}}
\preprint{DOE/ER/40762-322} \preprint{MU-PP\#05-006}

\begin{document}

\title{Atmospheric Neutrino Mixing and $b \rightarrow s$ Transitions: \\
Testing Lopsided SO(10) Flavor Structure in $B$ physics}
\author{Xiangdong Ji}
\affiliation{Department of Physics, University of Maryland, College
Park, Maryland 20742}\affiliation{Center for High-Energy Physics and
Institute of Theoretical Physics, \\ Peking University,  Beijing
100080, P. R. China}
\author{Yingchuan Li}
\affiliation{Department of Physics, University of Maryland, College Park,
Maryland 20742}
\author{Yue Zhang}\affiliation{Center for High-Energy
Physics and Institute of Theoretical Physics, \\ Peking University,
Beijing 100080, P. R. China}
\date{\today}

\begin{abstract}
We point out that the correlation between the large atmospheric
neutrino mixing and the $b \rightarrow s$ transition, if exists,
comes from the lopsided flavor structure in SO(10) grand unification
theory. We suggest testing the correlation by studying the deviation
of $S_{\phi K_S}$ and $S_{\eta' K_S}$ from the standard model
predictions along with the constraints from $\Delta M_{B_S}$ and
$b\rightarrow s \gamma$ in a realistic SO(10) model with lopsided
flavor structure. We find a specific correlation between $S_{\phi
K_S}$ and $S_{\eta' K_S}$ that is intrinsic to the lopsided
structure and discuss how to confirm or rule out this flavor pattern
by the more accurate measurements of these CP violation quantities
in $B$ decay.
\end{abstract}

\maketitle

\section{Introduction}

The standard model (SM), although has passed many precision
experimental tests, still has a number of outstanding problems which
beg for a more fundamental theory. In terms of phenomenology, there
are many places to test the more fundamental theory and, quite
often, it is not one piece of phenomenology but a specific pattern
of many pieces of phenomenology that fulfils this job.

Neutrino oscillation clearly indicates the beyond SM physics in that
it violates the accidental symmetry $L_{e,\mu,\tau}$ of SM lepton
flavor. Moreover, the tiny neutrino masses can be most naturally
explained by the seesaw mechanism \cite{seesaw}, indicating the
Majorana nature of neutrinos that leads to violation of another
accidental symmetry $B-L$ of SM at a very high scale close to the
grand unification theory (GUT) scale. For this reason, studying
neutrino oscillation in the GUT framework becomes particularly
interesting.


In SO(10) GUT, quarks and leptons are unified into a single 16
multiplet, yet experimental data shows that the 2-3 (generation)
mixing is dramatically different in the quark and lepton sectors.
While the quark mixing angle $\theta^{\rm CKM}_{23}$ is around 0.04,
the lepton mixing angle $\theta^{\rm PMNS}_{23}$ (atmospheric mixing
angle) is quite large and could be a maximum of $\pi/4$. This
difference can be rather naturally explained by the lopsidedness in
the 2-3 sector of mass matrices of down-type quarks and charged
leptons:
\begin{eqnarray}
D_{23} & \propto & \begin{pmatrix}  0 & \sigma+\epsilon/3 \\
-\epsilon/3 & 1
\end{pmatrix}, ~~
L_{23} \propto \begin{pmatrix}  0 & -\epsilon \\
\sigma+\epsilon & 1
\end{pmatrix},
\end{eqnarray}
where $\sigma$ is of order one and generates large mixing for
left-handed (LH) charged leptons and right-handed (RH) down-type
quarks, while $\epsilon$ is much smaller than one and generates tiny
mixing for RH charged leptons and LH down-type quarks. The
group-theoretical origin of this lopsidedness can be most
transparently seen in terms of the SU(5) multiplet--- it is the
charge conjugate of RH down-type quarks that sit together with LH
lepton doublet in the ${\bar 5}$ representation of SU(5). This
lopsided structure can also be generated from SO(10) group
structure.

Indeed, realistic SO(10) GUT models with lopsided mass matrices have
been constructed \cite{Albright:lopsided,JLM:lopsided} in the
literature. These different versions differ in many aspects but have
the common structure as shown in Eq. (1). These models fit very well
with all the fermion masses and mixings, including the neutrino
sector through the first type of sea-saw mechanism, as well as
producing the correct amount of baryon asymmetry through
leptogenesis \cite{leptogenesis}.

Given the successes of the SO(10) models with lopsided flavor
structure, we are motivated to address the following question: what
is the most characteristic feature of these models and where to test
it?

From the lopsided structure itself as shown in Eq. (1), it is clear
that the most characteristic feature of the models with lopsided
structure is the large RH 2-3 mixing of down-type quarks associated
with the large neutrino atmospheric mixing angle. The question is
where and how to see its signature in other places for testing
beyond SM physics. Clearly this RH mixing does not show up in the
quark CKM (Cabbibo-Kobayashi-Moskawa) matrix, which is, in fact, the
point of introducing lopsidedness to generate large lepton mixing
and small quark mixing simultaneously. However, in the supersymmetry
(SUSY) theory, this large RH mixing has the potential of generating
sizable off-diagonal elements of soft mass matrices of squarks
which, in turn, can be manifested in the flavor-changing neutral
current interaction of down-type quarks, namely, the $b \rightarrow
s$ transition.

The penguin dominated $b \rightarrow s$ transition has long been
regarded as the golden channel for probing new physics. Moreover, if
there are phases associated with the new physics contributing to
this transition, there could be new CP violations in B physics.
Within the SM, the indirect CP asymmetry parameter $S_{\phi K_S}$
and $S_{\eta' K_S}$ are essentially the same as that of
$B\rightarrow J/\psi K_S$: $S^{SM}_{\phi K_S}\simeq S^{SM}_{\eta'
K_S} \simeq S_{J/\psi K_S}={\rm Sin}2\beta = 0.685\pm0.032$.
However, the experimental values of $S_{\phi K_S}$ and $S_{\eta'
K_S}$ from BaBar and Belle \cite{cpexp} show large deviations from
the SM prediction:
\begin{eqnarray}
S^{\rm exp.}_{\phi K_S}&=&0.50\pm0.25^{+0.07}_{-0.04}  \,\,\,\,\,\,\,\,\,\,
({\rm BaBar}), \nonumber \\
&=&0.06\pm0.33\pm0.09  \,\,\,\,\, ({\rm Belle}), \nonumber \\
S^{\rm exp.}_{\eta' K_S}&=&0.27\pm0.14\pm0.03  \,\,\,\,\, ({\rm BaBar}),
\nonumber \\
&=&0.06\pm0.18\pm0.04  \,\,\,\,\,\, ({\rm Belle}),
\end{eqnarray}
with the average of $S^{\rm exp.}_{\phi K_S}=0.34\pm0.20$ and
$S^{\rm exp.}_{\eta' K_S}=0.41\pm0.11$, which display 1.7$\sigma$
and $2.5\sigma$ deviations from the SM predictions, respectively.
This significant discrepancy between SM prediction and experiment
data has generated tremendous amount of effort in searching beyond
SM physics.

Among these efforts, it has been pointed out in Ref.
\cite{murayama} that there could be correlation between the large
atmospheric mixing and the large $b\rightarrow s$ transition,
based on the connection between LH charged leptons and RH
down-type quarks in the framework of SO(10) GUT. However, it
should be noted that, in fact, this correlation depends
exclusively on the lopsided structure, which is the only way of
realizing the possible connection between LH charged leptons and
RH down-type quarks. Within other realistic SO(10) models
\cite{models} without lopsided structure, a set of parameters that
are of the same order are combined constructively and
destructively to give large atmospheric angle and small quark 2-3
mixing angle, respectively. Therefore, in these SO(10) models
\cite{models}, there is typically no large RH down-type quark
mixing associated with the large atmospheric mixing, and hence the
correlation between atmospheric mixing and the $b \rightarrow s$
transition is not realized.

Knowing the correlation between atmospheric neutrino mixing and
the $b \rightarrow s$ transition through the lopsided structure in
SO(10), we test this possibility by investigating the predictions
of CP conserving and CP violating observables associated with the
$b \rightarrow s$ transition in B physics from a particular
lopsided SUSY SO(10) model constructed by us \cite{JLM:lopsided}.
Our study shows that the $S_{\phi K_S}$ and $S_{\eta' K_S}$ could
indeed have large deviations from their SM values because of the
large $b \rightarrow s$ transition induced by the lopsided flavor
structure in the SUSY context. Moreover, we find a particular
pattern of correlation between $S_{\phi K_S}$ and $S_{\eta' K_S}$,
which makes this class of models with lopsided structure
distinguishable from other types of models. We expect, for
example, the similar result from the model in
\cite{Albright:lopsided}, because these two models are nearly the
same in the down-type quark sector.

This paper is organized as follows. Section II is devoted to the
calculation of flavor violation variables from the SO(10) model
described in \cite{JLM:lopsided}. In Sec. III, we present the
predictions of $S_{\phi K_S}$ and $S_{\eta' K_S}$, with the
constraint from $b \rightarrow s \gamma$ as well as the recent
measurement of $\Delta M_{B_S}$, in the model. Finally, we conclude
in Sec. IV.

\section{SUSY flavor violation parameters from the SO(10) GUT model}

The SUSY flavor violation and flavor-violating CP violation in the
quark sector are induced by the off-diagonal elements of squark
mass-squared matrices $m^2_{AB}$ with $A,B={L,R}$ indicating the
chirality. In the mass insertion approximation (MIA) approach, the
relevant parameters are $\delta_{AB}'$s, which are the $m^2_{AB}$
divided by the average squark mass-squared. We restrict ourselves to
studying the gluino contribution, which is believed to be the
dominant one due to enhancement by the large gauge coupling
$\alpha_S$ \cite{gluino}. In the gluino-induced contribution, the
relevant parameters for the $b \rightarrow s$ transition are the
$(\delta^d_{LL,RR,LR,RL})_{23}$ of down-type quark. We are going to
show how these parameters are calculated in the SO(10) GUT model in
this section.

With the universality condition imposed at the SUSY breaking scale
$M_*$, which we take as the Planck scale $M_{Pl.}=10^{19}{\rm GeV}$,
and with the magnitude of $\mu$-parameter fixed from the radiative
electroweak breaking, there are five SUSY parameters left: $(m_0,
m_{1/2}, A_0, {\rm tan} \beta, \phi_{\mu})$. ${\rm tan} \beta$ is
fixed to be 10 when our SO(10) model is constructed to fit fermion
masses \cite{JLM:lopsided}. The phase of $\mu$-parameter,
$\phi_{\mu}$,  is constrained from the electric dipole moment (EDM)
bounds, which, if assuming possible cancellation exists, restrict
$\phi_{\mu}$ to be within $\pm \pi/10$ from 0 or $\pi$
\cite{EDMsuppresion,EDMcancellation}. We take the $\phi_{\mu}$ to be
in the range of $(-\pi/10,\pi/10)$ for concreteness.
Furthermore, we assume $A_0=0$ at $M_*$ (see \cite{zeroAterm} for
justification). A nonzero $A_0$ should not bring any significant
change to the results since the alignment condition is assumed.
Finally, we set two soft masses $m_0$ and $m_{1/2}$, which are the
universal soft scalar mass and gaugino mass at $M_*$, respectively,
to be within 1 TeV.

The off-diagonal elements of squark mass-squared matrices are
generated from the renormalization group (RG) running between SUSY
breaking scale $M_*$ and the electroweak scale $M_{\rm EW}$. The GUT
symmetry breaking scale, $M_{\rm GUT}=2\times10^{16}{\rm GeV}$,
divides this running into two parts: above-$M_{\rm GUT}$ running and
below-$M_{\rm GUT}$ running.

In the following discussion, we will stick to the super-KM basis for
the squark fields, in which the neutral current quark-gaugino-squark
vertices are diagonal.

Below the $M_{\rm GUT}$, there are two Yukawa couplings in the
quark sector: ${\bar d} Y_d Q H_d$ and ${\bar u} Y_u Q H_u$. The
running of $(m^d)^2_{RR}$ is proportional to $Y_d Y^{\dagger}_d$
which is diagonal in the super-KM basis of RH down-type squarks.
Therefore no off-diagonal element of $(m^d)^2_{RR}$ should be
generated from the below-$M_{\rm GUT}$ running. Nevertheless, the
running of $(m^d)^2_{LL}$ involve both $Y^{\dagger}_d Y_d$ and
$Y^{\dagger}_u Y_u$. While the former is diagonal in the super-KM
basis of LH down-type squarks, the latter is not and could
generate the off-diagonal elements of $(m^d)^2_{LL}$. We have
\begin{equation}
(\delta^d_{LL})^{\rm below-GUT}_{ij}=-\frac{3}{8 \pi^2} (
Y^{\dagger}_u Y_u)_{ij} \ln(\frac{M_{GUT}}{M_{EW}})
\end{equation}
where the $Y_u$ is in the basis of SU(2) doublet $Q$ that $Y_d$ is
diagonal.

Above the $M_{\rm GUT}$, all the 16 fermions, including the RH
neutrino, are in the 16 spinor representation of SO(10). The soft
mass-squared $m^2_{16}$ is renormalized by the single renormalizable
operator $f_{33} 16_3 16_3 10_H$ in the model. As discussed in Ref.
\cite{aboveGUT}, the initial universal soft mass-squared
$(m^2_{16})|_{M_*}={\rm diag}(m^2_0,m^2_0,m^2_0)$ is not kept at
$M_{\rm GUT}$: $(m^2_{16})|_{M_{GUT}}={\rm
diag}(m^2_0,m^2_0,m^2_0-\Delta m^2)$. The change of 3-3 element
$\Delta m^2$ is due to the renormalization by the operator $f_{33}
16_3 16_3 10_H$:
\begin{equation}
\Delta m^2 =
\frac{60m^2_0}{16\pi^2}f^2_{33}\ln(\frac{M_*}{M_{GUT}}).
\end{equation}
The parameter $f_{33}$ is not completely fixed in the model and we
choose it to be 1/2, which is in the reasonable range.
This non-universal, diagonal, soft mass-squared matrix is in the GUT
basis. After being rotated to the super-KM basis, off-diagonal
elements of $(m^2)^d_{RR,LL}$ are generated:
\begin{equation}
(m^2)^d_{LL,RR}|_{\rm super-KM}=U^{\dagger}_{L,R}
(m^2_{16})|_{M_{GUT}} U_{L,R}
\end{equation}
where $U_{L,R}$ are the unitary transformation matrices that
diagonalize the down-type quark mass matrix $M^{\rm
digonal}_d=U^{\dagger}_R M_d U_L$. The $(\delta^d_{LL,RR})^{\rm
above-GUT}$ is obtained from $(m^2)^d_{LL,RR}|_{\rm super-KM}$
divided by the average of its diagonal elements. Finally, the
$\delta^d_{LL}$ is the sum of $(\delta^d_{LL})^{\rm below-GUT}$ and
$(\delta^d_{LL})^{\rm above-GUT}$, while the
$\delta^d_{RR}=(\delta^d_{RR})^{\rm above-GUT}$ is only from
above-GUT running.

The point mentioned in Sec. I that the correlation between
atmospheric neutrino mixing and $b \rightarrow s$ transition depends
exclusively on the lopsided structure can be explicitly seen here:
In the lopsided flavor structure, the 2-3 element of $M_d$ is large
and induces a large 2-3 rotation $\theta^R_{23}$ in $U_R$. This
large rotation in turn produces a large
$(\delta^d_{RR})^{above-GUT}_{23}$ as shown in Eq. (5). Finally, the
large off-diagonal squark masses can generate a large $b \rightarrow
s$ transition.

Although we set $A_0=0$ at $M_*$, it could be generated through
radiative corrections. For $(m^d)^2_{RL,LR}$, the running below
$M_{\rm GUT}$, being proportional to $Y_d$, only induces diagonal
elements in the super-KM basis. However, running from $M_*$ to
$M_{\rm GUT}$ does generate off-diagonal elements of $A_{LR,RL}$. In
the GUT basis
\begin{equation}
A^d_{RL}|_{\rm GUT}=c
\begin{pmatrix} \frac{63}{2}\eta & 45\delta & 45\delta'
\\ 45\delta & 0 & 45\sigma+61\epsilon/3 \\ 45\delta' & -61\epsilon/3 &
\frac{63}{2}
\end{pmatrix}M_D  ,
\end{equation}
where $c=\frac{1}{8\pi^2}g^2_{10} M_{1/2} \ln(\frac{M_*}{M_{GUT}})$
and $\eta$, $\delta$, $\delta'$, $\epsilon$, and $M_D$ are
parameters fixed in the model \cite{JLM:lopsided}. The
pre-coefficients 63/2, 45, and 61 are sums of Casimirs of SO(10)
representation involved in the operator $16_i 16_j 10_H$, $16_i 16_j
[\overline{16}_H \overline{16}'_H]_{10}$, and $16_i 16_j
[\overline{16}_H \overline{16}'_H]_{10} 45_H$, respectively. Again,
one simply applies $U_{R,L}$ on both sides of $A^d_{RL}$ to go to
the super-KM basis
\begin{equation}
A^d_{RL}|_{\rm super-KM}=U^{\dagger}_{R}A^d_{RL}|_{\rm GUT}U_{L},
\end{equation}
which, together with the diagonal $\mu$ term contribution, gives the
full $(m^d)^2_{RL}$: $(m^d)^2_{RL}=A^d_{RL}|_{\rm super-KM}-\mu {\rm
tan} \beta~ {\rm diag}(m_d,m_s,m_b)$. Finally, $\delta^d_{RL,LR}$ is
obtained as $\delta^d_{RL,LR}=(m^d)^2_{RL,LR}/m^2_0$.

To see the characteristic feature of the lopsided structure, it is
instructive to look at the size of all the $\delta$'s. Taking
$m_0=300~{\rm GeV}$, $m_{1/2}=500~{\rm GeV}$, and $\phi_\mu=\pi/10$
as an example, we have $(\delta^d_{RR})_{23}=0.28 e^{-0.05 i}$,
$(\delta^d_{LL})_{23}=0.0028 e^{-0.07 i}$,
$(\delta^d_{LR})_{23}=0.00003 e^{-0.05 i}$,
$(\delta^d_{RL})_{23}=-0.0009 e^{-0.05 i}$. Obviously, the
$(\delta^d_{RR})_{23}$ is of several orders of magnitude larger than
all the other $\delta$'s. Moreover, giving this large
$(\delta^d_{RR})_{23}$, a large effective $(\delta^d_{RL})^{\rm
eff}_{23}$ is generated from the double mass insertion:
$(\delta^d_{RL})^{\rm eff}_{23} = (\delta^d_{RL})_{23} +
(\delta^d_{RR})_{23} \mu {\rm tan}\beta m_b/m^2_0$, which is $0.064
e^{-0.38 i}$ for the same set of parameters. The effective
$(\delta^d_{LR})^{\rm eff}_{23}$ still remains small in the case of
double mass insertion due to the smallness of
$(\delta^d_{LL})_{23}$. As a result, the lopsided model predicts
large $(\delta^d_{RR})_{23}$ and $(\delta^d_{RL})^{\rm eff}_{23}$,
and small $(\delta^d_{LL})_{23}$ and $(\delta^d_{LR})^{\rm
eff}_{23}$.

\section{Flavor-Changing Neutral Current Effects in B Mesons}

The most general effective Hamiltonians $H^{\Delta B=1}_{eff}$ and
$H^{\Delta B=2}_{eff}$ for the non-leptonic $\Delta B=1$ and $\Delta
B=2$ processes are
\begin{eqnarray}
H^{\Delta B=1}_{\rm eff} &=&
\frac{G_F}{\sqrt{2}}\sum_{i=1\sim10,7\gamma,8g} \{ C_i(\mu) Q_i(\mu)
+
\tilde{C}_i(\mu) \tilde{Q}_i(\mu) \}  \nonumber \\
H^{\Delta B=2}_{\rm eff} &=& \sum^5_{i=1} C_i(\mu) Q_i(\mu) +
\sum^3_{i=1} \tilde{C}_i(\mu) \tilde{Q}_i(\mu) ,
\end{eqnarray}
where $C_i(\mu)$, $\tilde{C}_i(\mu)$ and $Q_i(\mu)$,
$\tilde{Q}_i(\mu)$ are the Wilson coefficients and the local
operators (not same in both hamiltonians), respectively. All the
relevant contributions of high energy physics above $W$ mass,
including the SUSY particle contribution, enter the Wilson
coefficients at $\mu=m_W$: $C(m_W)$ and $\tilde{C}(m_W)$. The matrix
elements of local operators are, however, obtained at the energy
scale of bottom quark mass $m_b$. Therefore, one needs to obtain the
Wilson coefficients at low energy by solving the renormalization
group equations of QCD and QED in the SM:
\begin{equation}
C_i(m_b)=\sum_j \hat{U}(m_b, m_W)C_j(m_W) \ ,
\end{equation}
where the evolution matrix $\hat{U}(m_b, m_W)$ for $\Delta B=1$
and $\Delta B=2$ Wilson coefficients can be found in Ref.
\cite{evolution1} and Ref. \cite{evolution2}, respectively.

The SM and SUSY contributions to Wilson coefficients can be found in
Refs. \cite{Gabbiani,Gabrielli}. It is worth noting that the SUSY
contribution depends on the squark mass $m_{\tilde{q}}$ and gluino
mass $m_{\tilde{g}}$, which are larger than the universal soft
scalar mass $m_0$ and gaugino mass $m_{1/2}$ due to the RG running.
We use the matrix elements of local operators evaluated in QCD
factorization (QCDF), developed in Ref. \cite{QCDF}, which makes the
strong phase calculable, yet introduces undetermined parameters
$\rho_{H,A}$ and phases $\phi_{H,A}$.

To make prediction of $S_{\phi K_S}$, we first impose constraints on
the parameter space by requiring the prediction of branching ratio
and CP asymmetry of $b \rightarrow s \gamma$ and $\Delta M_{B_S}$ to
be within the experimental bounds.

The gluino contribution to the branching ratio $b \rightarrow s
\gamma$ is \cite{Gabbiani}
\begin{eqnarray}
BR(b \rightarrow s \gamma)_{\tilde{g}} &=& \frac{\alpha^2_s
\alpha}{81 \pi^2 m^4_{\tilde q}} \left\{ |m_b
M_3(x)(\delta^d_{LL})_{23} \right. \\
&&  \left. + m_{\tilde{g}} M_1(x)(\delta^d_{LR})_{23} |^2 + L
\leftrightarrow R \right\}, \nonumber
\end{eqnarray}
where the loop functions $M_1(x)$ and $M_3(x)$ with
$x=m^2_{\tilde{q}}/m^2_{\tilde{g}}$ can be found in Ref.
\cite{Gabbiani}. As discussed in Ref. \cite{Gabbiani}, The
experimental bound and the SM uncertainty together require that the
gluino contribution $BR(b \rightarrow s \gamma)_{\tilde{g}} < 4
\times 10^{-4}$. The bound on the CP asymmetry $A^{CP}_{b
\rightarrow s \gamma}$ plays no significant role in constraining the
parameter space. Therefore we neglect its discussion here, although
we have included it in the calculation in the same way as in Ref.
\cite{Gabrielli}.


The D0 and CDF Collaborations \cite{deltaM} have reported new
results for $\Delta M_{B_S}$:
\begin{eqnarray}
17~{\rm ps}^{-1}<\Delta M_{B_S}<21~ {\rm ps}^{-1} && ({\rm D0}),  \nonumber
\\
\Delta M_{B_S}=17.33^{+0.42}_{-0.21}\pm0.07 ~{\rm ps}^{-1}  && ({\rm
CDF}),
\end{eqnarray}
while the best fit value in SM is $\Delta M_{B_S}=17.5~{\rm
ps}^{-1}$. This imposes the constraint $|R_M| \equiv
|M^{SUSY}_{12}/M^{SM}_{12}|\leq 4/17$, where $M_{12}=\langle B^0_s|
H^{\Delta B=2}_{\rm eff} | \overline{B}^0_s \rangle$. One should
notice that this bound remains valid if one considers the
uncertainty in the SM value and assumes $\Delta M_{B_S}=21~{\rm }
ps^{-1}$ \cite{khalil}.

\begin{figure}[t]
\begin{center}
\mbox{
\begin{picture}(0,100)(60,0)

\put(-30,-20){\insertfig{6.5}{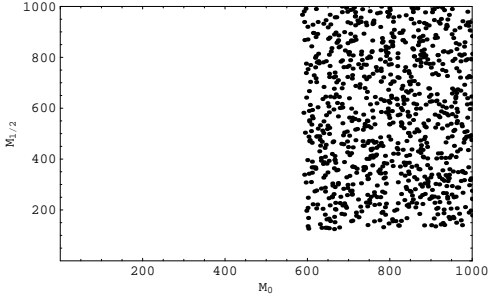}}

\end{picture}
}
\end{center}
\caption{\label{fig1} Scatter plot of the parameter space of $m_0$
and $m_{1/2}$ with the constraints from $BR(b \rightarrow s
\gamma)$, $A^{CP}_{b \rightarrow s \gamma}$ and $\Delta M_{B_S}$.}
\end{figure}

\begin{figure}[t]
\begin{center}
\mbox{
\begin{picture}(0,220)(60,0)

\put(-30,100){\insertfig{6.5}{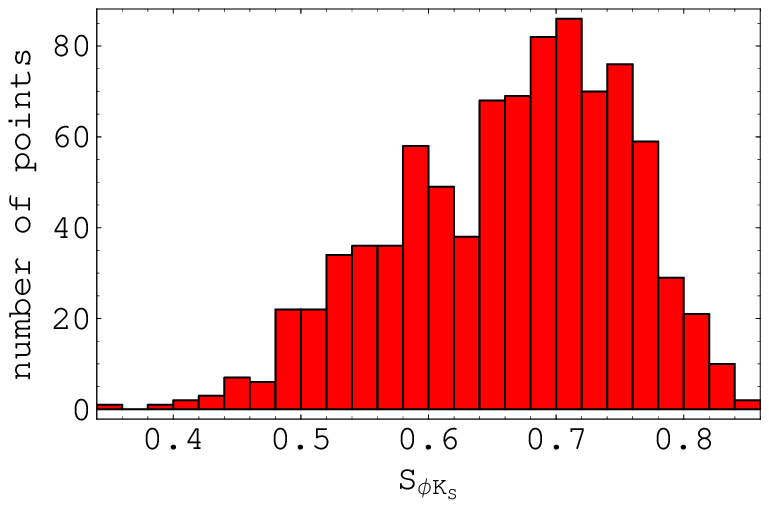}}

\put(-30,-20){\insertfig{6.5}{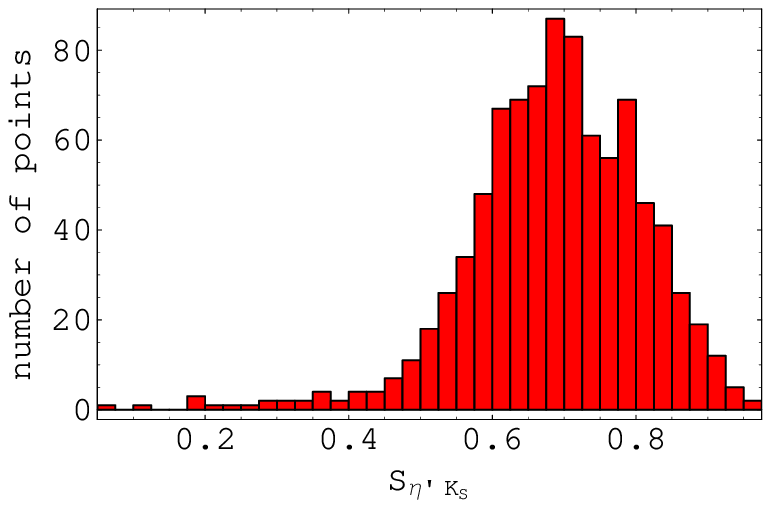}}

\end{picture}
}
\end{center}
\caption{\label{fig1} Predictions of $S_{\phi K_S}$ and $S_{\eta'
K_S}$ corresponding to the points in Fig. 1.}
\end{figure}

\begin{figure}[t]
\begin{center}
\mbox{
\begin{picture}(0,65)(60,0)

\put(-65,-20){\insertfig{4.4}{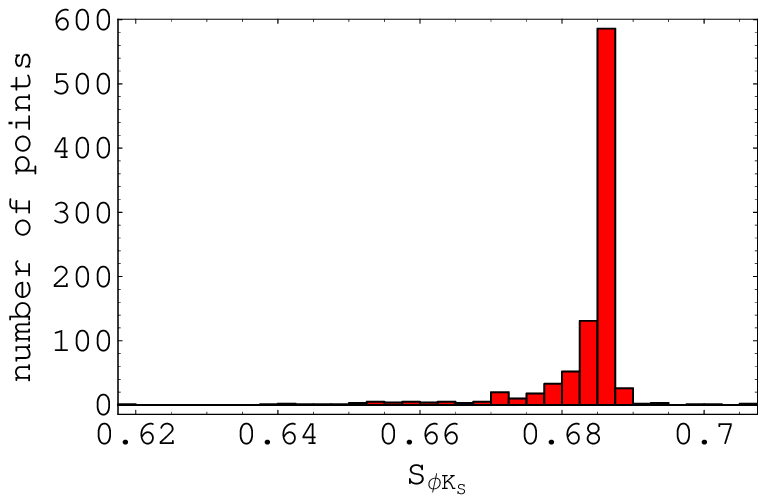}}

\put(65,-20){\insertfig{4.4}{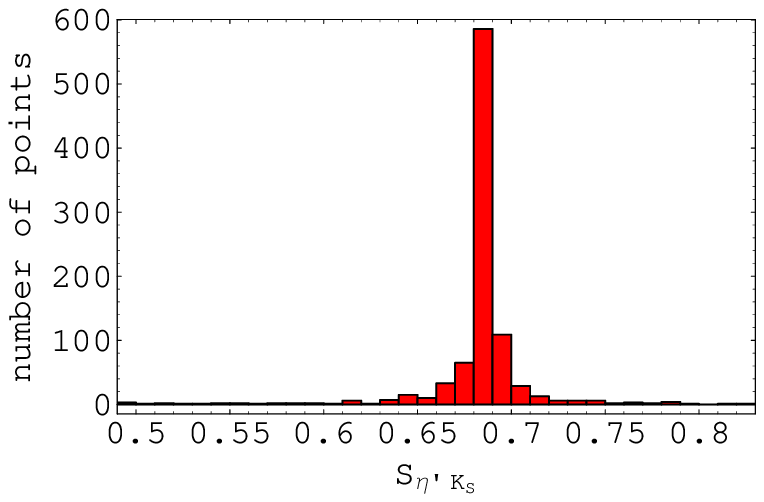}}

\end{picture}
}
\end{center}
\caption{\label{fig1} Predictions of $S_{\phi K_S}$ and $S_{\eta'
K_S}$ in the case that there is no large $(\delta^d_{RR})_{23}$
and its induced $(\delta^d_{RL})_{23}$.}
\end{figure}

The decay amplitudes of $B_d \rightarrow \phi K_s$ are given by
\cite{Gabrielli}
\begin{eqnarray}
A_{B_d \rightarrow \phi K_s} &= &-i \frac{G_F}{\sqrt{2}} m^2_{B_d}
F^{B_d \rightarrow K_s}_+ f_{\phi} \nonumber \\
&&\times \sum_{i=1\sim10,7\gamma,8g}H_i(\phi)(C_i + \tilde{C}_i)
\end{eqnarray}
where $f_{\phi}=0.233{\rm GeV}$ and $F^{B_d \rightarrow
K_s}_+=0.35$ is the transition form factor evaluated at transfered
momentum of order of $m_{\phi}$. The $H_i$'s are dependent on QCDF
parameters $\rho_{H,A}$ and $\phi_{H,A}$ in such a way as given in
Ref. \cite{Gabrielli}.

The SUSY contribution modifies the CP asymmetry as
\begin{equation}
S_{\phi K_S}={\rm sin}2\beta + 2 {\rm cos}2\beta {\rm sin}
\theta_{\phi} {\rm cos}\delta_{\phi} R_{\phi} + O(R^2_{\phi})
\end{equation}
where $R_{\phi}$, $\theta_{\phi}$, and $\delta_{\phi}$ are defined
in the ratio
\begin{equation}
A^{SUSY}_{B_d \rightarrow \phi K_S}/A^{SM}_{B_d \rightarrow \phi
K_S} \equiv R_{\phi} e^{i \theta_{\phi}} e^{i \delta_{\phi}}
\end{equation}
where the $R_{\phi}$ is the absolute value of the ratio, the
$\theta_{\phi}$ is the SUSY CP violating weak phase which depends
on the phases in $\delta$s, and the
$\delta_{\phi}=\delta^{SM}_{\phi}-\delta^{SUSY}_{\phi}$ is the CP
conserving strong phase that depends on $\phi_{H,A}$.

We deal with $S_{\eta' K_S}$ in the similar way as we deal with
$S_{\phi K_S}$, with the relevant coefficients also found in Ref.
\cite{Gabrielli}. One thing worth noting is that due to the fact
that, contrary to the $B \rightarrow \phi K$ transition, the
initial and final states in $B \rightarrow \eta' K$ transition
have opposite parity and therefore $\langle \eta' K | Q_i | B
\rangle=-\langle \eta' K | \tilde{Q}_i | B \rangle$, $C_i$ and
$\tilde{C}_i$ appear in such combinations as $C_i - \tilde{C}_i$
in the amplitude $A_{B_d \rightarrow \eta' K_s}$ instead of $C_i +
\tilde{C}_i$ in $A_{B_d \rightarrow \phi K_s}$ as shown in Eq.
(12). Since $C_{7\gamma,8g}$ depends on $(\delta^d_{LR})_{23}$,
while $\tilde{C}_{7\gamma,8g}$ depends on $(\delta^d_{RL})_{23}$,
this difference makes the correlation between $S_{\phi K_S}$ and
$S_{\eta' K_S}$ in the case with large $(\delta^d_{LR})_{23}$
different from the case with large $(\delta^d_{RL})_{23}$. In
fact, as shown in the general analysis in Ref. \cite{Gabrielli},
the deviations of $S_{\phi K_S}$ and $S_{\eta' K_S}$ from the SM
value are in the same direction if $(\delta^d_{LR})_{23}$ is large
and in the opposite direction if $(\delta^d_{RL})_{23}$ is large.
This turns out to be important in the following discussion of the
correlation of $S_{\phi K_S}$ and $S_{\eta' K_S}$ predictions from
the lopsided model.

Besides the SUSY parameters $(m_0, m_{1/2}, A_0, {\rm tan} \beta,
\phi_{\mu})$ which are discussed in Sec. II, the undetermined
parameters $\rho_{H,A}$ are constrained by $BR(B_d \rightarrow
\phi K_S)$ to be within $\rho_{H,A} \leq 2$ \cite{Gabrielli}, and
the strong phase $\phi_{H,A}$ is not constrained.

By scanning over the allowed ranges of undetermined parameters
$(m_0, m_{1/2},\phi_{\mu},\rho_{H,A},\phi_{H,A})$ and imposing the
bound of $\Delta M_{B_S}$, as well as $BR(b \rightarrow s \gamma)$
and $A^{CP}_{b \rightarrow s \gamma}$, we find the allowed
$(m_0,m_{1/2})$ shown in Fig. 1: There is a large parameter space of
$m_0, m_{1/2}$ satisfying the bound.

The corresponding predictions of $S_{\phi K_S}$ and $S_{\eta'
K_S}$ are shown in Fig. 2, from which we see that the large
$(\delta^d_{RR})_{23}$ does push the $S_{\phi K_S}$ and $S_{\eta'
K_S}$ off their SM value $0.685$. For the purpose of comparison,
we set by hand the $(\delta^d_{RR})_{23}$ to be of the size of
$(\delta^d_{LL})_{23}$ , which would be the case without lopsided
structure, and present the corresponding prediction of $S_{\phi
K_S}$ and $S_{\eta' K_S}$ in Fig. 3, which, together with Fig. 2,
shows clearly that the large deviation of $S_{\phi K_S}$ and
$S_{\eta' K_S}$ from their SM values are exclusively due to the
large $(\delta^d_{RR})_{23}$ from the lopsided structure.


\begin{figure}[t]
\begin{center}
\mbox{
\begin{picture}(0,120)(60,0)

\put(-30,-10){\insertfig{6.5}{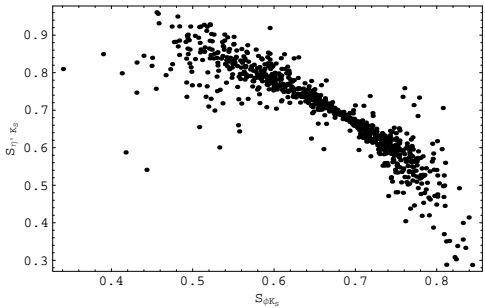}}

\end{picture}
}
\end{center}
\caption{\label{fig1} Scatter plot of predictions of $S_{\phi
K_S}$ and $S_{\eta' K_S}$ corresponding to the points in Fig. 1.}
\end{figure}

While Fig. 2 shows that the lopsided structure may explain the
anomalies of both $S_{\phi K_S}$ and $S_{\eta' K_S}$, the
correlation between the predictions of these two quantities, shown
in Fig. 4, indicate an interesting pattern: the large
$(\delta^d_{RL})_{23}$ push $S_{\phi K_S}$ and $S_{\eta' K_S}$ in
opposite directions. For points that $S_{\phi K_S}$ goes small, the
$S_{\eta' K_S}$ goes larger, and vice versa. As discussed above,
this specific pattern is intrinsic to the large
$(\delta^d_{RR})_{23}$, which induces large $(\delta^d_{RL})_{23}$
yet leaves $(\delta^d_{LR})_{23}$ small. Therefore, it is tightly
associated with the lopsided flavor structure. This specific pattern
of correlation between $S_{\phi K_S}$ and $S_{\eta' K_S}$ means that
the lopsided flavor structure cannot be responsible for both
anomalies simultaneously. If future experiments confirm that
$S_{\phi K_S}$ and $S_{\eta' K_S}$ are indeed both significantly
smaller than the SM values, the lopsided SO(10) model is ruled out
unless one assumes that SUSY parameters are such that large
$(\delta^d_{RR})_{23}$ from the lopsided structure makes no
significant contribution to the $b \rightarrow s$ transition and the
$S_{\phi K_S}$ and $S_{\eta' K_S}$ anomalies are from other beyond
SM physics sources. On the other hand, if future experiments show
the deviation of $S_{\phi K_S}$ and $S_{\eta' K_S}$ from their SM
values in opposite directions, it would be a strong evidence for the
lopsided flavor structure.

\section{conclusion}

In this paper, we pointed out that a possible correlation between
large atmospheric neutrino mixing and the $b \rightarrow s$
transition, first discussed in Ref. \cite{murayama}, in fact,
depends exclusively on the lopsided SO(10) structure. We studied the
prediction of $S_{\phi K_S}$ and $S_{\eta' K_S}$ from a realistic
SO(10) model with lopsided flavor structure with the constraints
from $\Delta M_{B_S}$, $b\rightarrow s \gamma$ applied. We found
that both quantities can show significant deviations from their SM
values due to the lopsided flavor structure, but with a specific
type of correlation. We discussed that the specific correlation of
the two quantities can be used to test the flavor structure by
future experiments.

X. Ji and Y. Li are partially supported by the U. S. Department of
Energy via grant DE-FG02-93ER-40762 and by National Natural
Science Foundation of China (NSFC), and Y. Zhang is supported by
the NSFC grants 10421503 and 10625521. Y. Li thanks R. N.
Mohapatra, M. Ramsey-Musolf and P. Rastogi for helpful discussions
and Z. Z. Xing for his hospitality at High Energy Institute of
Physics, Beijing where part of this research was completed.

\appendix

\end{document}